\documentclass[conference]{IEEEtran}

\ifCLASSINFOpdf
\else
\fi

\usepackage{mdwmath}
\usepackage{mdwtab}

\usepackage{amsmath}
\usepackage{ amssymb}
\usepackage{amsmath}
\usepackage{amsfonts}
\usepackage{amssymb}
\usepackage{graphicx}
\usepackage[hang,small,bf]{caption}

\usepackage{subfigure}
\usepackage{etoolbox}

\usepackage{hyperref}
 \hypersetup{bookmarks=false,colorlinks=false}

\begin{document}
\title{Optimal Threshold Scheduler for Cellular Networks }

\author{\IEEEauthorblockN{Sanket Kamthe}
\IEEEauthorblockA{Fachbereich Elektrotechnik und Informationstechnik\\
TU Darmstadt\\
Merck str. 25,
64283 Darmstadt \\
Email: sanket.kamthe@stud.tu-darmstadt.de}
\and
\IEEEauthorblockN{Smriti Gopinath}
\IEEEauthorblockA{Fachbereich Elektrotechnik und Informationstechnik\\
TU Darmstadt\\
Merck str. 25,
64283 Darmstadt \\
Email: smriti.gopinath@stud.tu-darmstadt.de}
}


%Telephone: (800) 555--1212, Fax: (888) 555--1212}


\maketitle

\begin{abstract}
The conventional wireless schedulers of Unicast and Multicast either fulfill multiuser diversity or broadcast gain, but not both together. To achieve optimal system throughput we need a scheduler that exploits both the multiuser diversity gain and the Multicasting gain simultaneously. We first propose a new median-threshold scheduler that selects all users having instantaneous SNR above the median value for transmission. The system rate equation for the proposed scheduler is also derived. We then optimize the median-threshold so that it performs well over an entire SNR range. With the help of simulation results we compare performance of the proposed scheduler with schedulers like Unicast, Multicast and other Opportunistic Multicast Schedulers (OMS) which are the best schedulers in terms of throughput and show that the proposed optimized threshold scheme outperforms all of them. 
\end{abstract}


% \fi
\IEEEpeerreviewmaketitle

\section{Introduction}
 \IEEEPARstart Multimedia applications such as video conferencing,streaming etc.require more network resources than voice data applications. Presently, data to be transmitted must be duplicated if there are multiple users who want to receive the same data. Such an approach is called Unicast scheduling in which the user with the best instantaneous channel realization (highest supportable data rate) is served at a time.  This is from the phenomenon of multiuser diversity- for many users that have independent channel fading characteristics, at a given time there is high probability that one of the users has a strong channel. As the ever increasing demand for multimedia data will exceed the available resources Unicast leads to waste of radio resource and also delays while transmitting as each user has to wait for its turn.\\

On the other hand in Multicast scheduling the data is sent at the lowest data rate supported by the user with the worst channel realization. The main benefit of Multicast is bandwidth saving as it reduces redundant transmission of the same information. Multicasting gain is achieved when a single wireless transmission intended to the user with the lowest supportable rate can also be decoded by other users with better channel realizations.But on the flipside the rates are limited to the worst case even for users who have better channel conditions which brings down the system throughput.\\
\\
We first explain the explain the system model in section 2. In section 3, we give the details of our proposed median-threshold scheduling algorithm and compare it with existing schemes. The optimized version of the median threshold scheduler is presented in section 4. Simulation results are shown in section 5, and finally we conclude our paper in section 6.

\section{System Model}

Here we consider a down link Multicast scene of a single cell wireless communication network where one base station (BS) transmits packet in time slots of fixed duration to N users.The focus is on homogeneous networks where all users are equidistant from BS and experience independent and identically distributed (i.i.d.) channel conditions.

For the channel model we consider a frequency flat Raleigh fading channel. The distribution of instantaneous SNR is given by

\begin{equation} \label{equ: Exp variable}
f(x) =
\begin{cases}
\frac{1}{\overline{\gamma }}\exp \left( \frac{-x}{\overline{\gamma }} \right) & \text{if } x\ge {0},\\
0 & \text{Otherwise } ,\\

\end{cases}
\end{equation}

Where, $\overline{\gamma}$ is average SNR 

\label{1}

We consider a system based on Shannon\textquoteright s rate formula for maximum capacity that assumes a perfect error correcting system. The maximum bitrate of user 
\[{{R}_{\max }}(bps)=W{{\log }_{2}}\left( 1+\gamma  \right)\]
Where $\gamma$ is the average received SNR and $W$ is system Bandwidth. We assume system bandwidth $W=1$ from hereon in this paper.

\section{Median Threshold Scheduler}

We propose a new scheme/Scheduler which can be described by following psuedocode\\
\begin{figure}[h!]
\begin{center}\small
\begin{tabular}{|l|}\hline
\begin{minipage}{0.9\hsize}
\vspace{3mm}

\textbf{Intialization:}\\
Set ${{\gamma }_{th}}={{\gamma }_{Median}}$ \\
\textbf{Recursion:} (At each TTI)\\
Select the users passing threshold ${{\gamma }_{th}}$ \\
Transmit with $\inf \left( \gamma >{{\gamma }_{th}} \right)$ \\
If none of the users pass the threshold \\
Transmit with $\max \left( \gamma  \right)$\\ 
\vspace{1mm}
\end{minipage} \\ \hline
\end{tabular}
\end{center}
\caption{System Pseudo code}
\label{Pseudo-code}

\end{figure}
\subsection{Comparison with other Schedulers}

This scheduler exploits multiuser diversity provided by Knopp and Humblet model \cite{Knoop_and_Humblet}. The current model exploits this diversity by selecting users which have better realizations and discarding the ones that have poor realizations. Deciding the number of users to be selected is an important criteria that has been studied extensively. In the following we briefly analyze why our proposed scheme performs better than conventional schemes.

\subsubsection{Unicast or Max SNR or Best user Scheduling}

Unicast utilizes multiuser diversity by always choosing user with best channel realization. Such a system increases throughput when the user  with the best realization is selected for transmission. It fails to capitalize in scenarios where more than one user has good instantaneous values. Proposed scheduler does not suffer from such a drawback as it serves all the users passing threshold thereby selecting all users having high $\gamma$  values. It can be seen from simulation results that proposed scheme always outperforms Unicast Scheduling at all SNR regimes.

\subsubsection{Multicast  or Broadcast or Worst user Scheduling}

Multicast Scheduler transmits data to all the users in each time slot. This significantly increases the system throughput in High SNR regime. However, Multicast scheduler has to serve at the rate of the worst channel realization, and hence, gets severely affected by the poor $\gamma 's$ . Proposed scheduler overcomes this problem by discarding users which have lower$\gamma 's$ and this results in better performance compared to Multicast in low SNR regimes.

\subsubsection{Median User Scheduler}

  To overcome problems of Unicast and Multicast Median User Scheduler has been proposed by Gopala et al. \cite{Praveenkumar_throuputdelay,Praveen_Kumar}
Median User scheduler always selects N/2 users for scheduling. This scheme always outperforms Unicast and gives better performance than Multicast upto 20 dB . The proposed scheduler performs better than median user scheduler under all SNR regimes. This can be explained by observing that median scheduler always selects same number of users whereas proposed scheduler selects all users having what we consider to be a good channel condition. The number of users served \textquoteleft m\textquoteright  is itself a random number with binomial distribution. It is interesting to note that proposed scheduler also serves N/2 users on average.

\subsubsection{Opportunistic Multicast Scheduling }

\begin{figure}[h!]
  \vspace{-10pt}
  \begin{center}
    \includegraphics[width=0.5\textwidth]{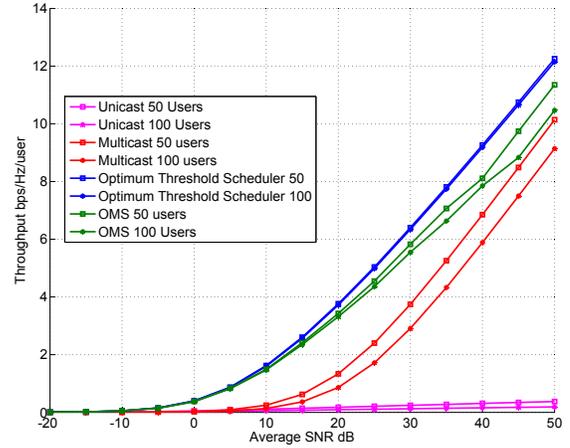}
  \end{center}
  \vspace{-10pt}
  \caption{System throughput comparison Optimised Threshold Scheduler}\label{Comparison of schedulers}
  %\vspace{-100pt}
\end{figure}

An opportunistic multicast scheduler, that adjusts the user selection ratio threshold according to the instantaneous channel condition of users has been investigated in \cite{OMS}, in which it has been shown to outperform unicast, multicast and the median-user. The proposed scheme is shown to perform better than the existing opportunistic multicasting schemes in fig \ref{Comparison of schedulers}

\subsection{Median Threshold Scheduler Rate}

As it is clear from above discussion selecting ${{\gamma }_{th}}$ is critical to system performance. Instead of selecting arbitrary ${{\gamma }_{th}}$we select ${{\gamma }_{th}}={{\gamma }_{Median}}$ for the proposed ''Median Threshold Scheduler.''\\
Important observations when ${{\gamma }_{th}}={{\gamma }_{Median}}$\\
\begin{enumerate}
\item Each user has probability of being served p=1/2. Which implies each user is served in 1/2 number of time slots. This is from the definition of the median.
\item The threshold is selected based on an quantile function e.g. median it is a constant for a given  $\overline{\gamma }$. By definition of median each user has the same probability P=1/2 of being selected for transmission. This is a significant result as this makes,for a sufficiently large number of users, the system rate independent of number of users as will be shown later
\item As noted above unlike schedulers proposed till now number of users selected is a binomial random variable and not a predetermined number.
\item One can establish a lower bound on the system throughput given by 
\[{{R}_{low }}=\frac{1}{2}\times \left( {{\log }_{2}}(1+\ln (2)\overline{\gamma }) \right)\times B\]
This result can be obtained by substituting p=1/2 in final result derived in 
 Appendix \ref{Appendix: lower bound}
\end{enumerate}

Now we derive an expression for Rate of the system :
Assuming frequency flat Raleigh fading channel model PDF of system is given by \cite{Knoop_and_Humblet} 

\begin{equation}
f(x) =
\begin{cases}
\frac{1}{\overline{\gamma }}\exp \left( \frac{-x}{\overline{\gamma }} \right) & \text{if } x\ge {{\gamma }_{th}},\\
0 & \text{Otherwise } ,\\

\end{cases}
\end{equation}

As we transmit to user only when $\gamma >{{\gamma }_{th}}$ . We can find PDF of $\gamma$ by applying Baye’s rule
 
\[P(\gamma |\gamma >{{\gamma }_{th}})=P(\gamma ,\gamma >{{\gamma }_{th}})/P(\gamma >{{\gamma }_{th}})\] 
As ${{\gamma }_{th}}={{\gamma }_{Median}}$  we have by definition of median \\
	\[P(\gamma >{{\gamma }_{th}})={1}/{2}\;\]
 
substituting

\begin{equation}
p(\gamma |\gamma >{{\gamma }_{th}}) =
\begin{cases}
 2f(x) & \text{if } x\ge {{\gamma }_{th}},\\
0 & \text{Otherwise } ,\\

\end{cases}
\end{equation}

Integrating the same with limits from ${{\gamma }_{Median}}$ to $x$ we obtain CDF as\\ 
\[F(x)=1-2\exp \left( \frac{-x}{\overline{\gamma }} \right)\] \\
System transfers at minimum rate for all users passing threshold. Hence, under the assumption that all users are i.i.d variables with same average $\overline{\gamma }$ , minimum order statistic can be obtained by using survivor function \\
\[\widetilde{F}(x)={{\left[ 1-F(x) \right]}^{N}}\]\\ 
Substituting value of $F\left( x \right)$   \\
	\[\widetilde{F}(x)={{\left[ 2\exp \left( \frac{-x}{\overline{\gamma }} \right) \right]}^{m}}\]
 
where m=number of users above threshold
As the probability of user passing a threshold is an independent experiment, probability of \textquoteleft m\textquoteright  users passing threshold is a Bernoulli trial with success rate p=1/2.
Hence, the average survivor function of the system is given by\\
 \[\sum\limits_{m=1}^{N}{\left( P(m) \right)\left[ {{\left( \widetilde{F}(x) \right)}^{m}} \right]}\] \\
Where, P (m) is Probability Mass Function PMF of m users passing the threshold with probability of success p
\[P(m)=\left( \begin{matrix}
   N  \\
   m  \\
\end{matrix} \right){{p}^{m}}{{(1-p)}^{N-m}}\]	
Average value of random variable is given by 
\[\text{E} \left[ X \right]=\int\limits_{0}^{\infty }{\left( 1-F\left( x \right) \right)}dx\]
Hence, the average rate of the system is given by

%	\[{{R}_{avg}}=\int\limits_{{{\gamma }_{th}}}^{\infty }{\left( \sum\limits_{m=1}^{N}{\left( P(m) \right)\left[ {{\left( \text{ }\widetilde{F}(x) \right)}^{m}} \right]} \right)}\text{ }dx\]
\[{{R}_{avg}}=\int\limits_{{{R}_{th}}}^{\infty }{\left( \sum\limits_{m=1}^{N}{\left( P(m) \right)\left[ {{\left( \widetilde F({{2}^{x}}-1) \right)}^{m}} \right]} \right)}\text{ }dx\]
%\[{{R}_{avg}}=\int\limits_{{{\gamma }_{th}}}^{\infty }{\left( \sum\limits_{m=1}^{N}{\left( P(m) \right)\left[ {{\left( \text{  }\widetilde F(x) \right)}^{m}} \right]} \right)}\text{ }\left( \text{lo}{{\text{g}}_{\text{2}}}\left( \text{1+x} \right) \right)dx\]

Where, ${{R}_{th}}$ is given by \ref{equ:system lim}
It should be noted that in above expression we did not substitute p=1/2 to simplify expression. This is done in order to avoid repetition of same formula when we generalize the system in next section with optimal p*.\\

\section{Optimal Threshold Scheduler }

In Median Threshold Scheduler we selected ${{\gamma }_{th}}$ as Median of distribution. Is this threshold optimal ? We observe that that threshold value plays a crucial role in system throughput. To find value of optimum threshold we will use system lower bound as cost function for following reasons\\
\begin{enumerate}
\item It’s a concave function in p
\item Its independent of number of users and depends only on average SNR
\item A closed form solution can be obtained (shown in Appendix \ref{Appendix:optimzation})
\item Maximizing the minimum rate gives robustness 
\end{enumerate}

As we are using system lower bound as cost function rather than actual system rate for optimization, we  analyze the variation in peak values due to this approximation. It can be shown that this approximation does not significantly change peak values. 

%
%\begin{figure}[h]
%\begin{minipage}{0.45\linewidth}
%%\begin{tabular}{l}
%%\centering
%\begin{flushleft}
%\vspace{0pt}
%\includegraphics[width=1.2\textwidth]{./fig_OptimizationforNthreshold.pdf}
%\vspace{-80pt}
%\caption*{Threshold calculations}
%\end{flushleft}
%%\end{tabular}
%\end{minipage}
%\hspace{0.5cm}
%\begin{minipage}{0.45\linewidth}
%%\begin{tabular}{r}
%\begin{flushright}
%\vspace{0pt}
%\includegraphics[width=1.2\textwidth]{./fig_OptimizationforNrates.pdf}
%\vspace{-80pt}
%\caption*{Comparison of rates}
%\end{flushright}
%%\end{tabular}
%\end{minipage}
%\caption[Approximating optimization cost function]{Optimum Threshold: Effect of approximating optimization cost function}
%\label{Fig:Cost function}
%\end{figure}

Using system lower bound ( equ. \ref{equ:system lim} ) as  cost function we set up optimization problem as follows.   
\[p*=  \operatorname*{arg\,max}_p \quad [ \left( 1-p \right) {{\log }_{2}}\left( 1-\ln \left( 1-p \right)\times \overline{\gamma } \right)] \] 
We obtained a closed form solution to this optimization problem as  ( \ref{equ: optimal p} )

\[p*=\frac{{{e}^{\frac{{{e}^{W(\overline{\gamma })}}}{\overline{\gamma }}}}-\sqrt[\overline{\gamma }]{e}}{{{e}^{\frac{{{e}^{W(\overline{\gamma })}}}{\overline{\gamma }}}}}\]  
Where $W\left( x \right)$is Lambert function \cite{Lambert_W} defined as principal root of equation $Z=W{{e}^{W}}$  

This result is derived in detail in Appendix \ref{Appendix:optimzation}

\begin{figure}[h!]
  %\vspace{-20pt}
  \begin{center}
    \includegraphics[width=0.5\textwidth]{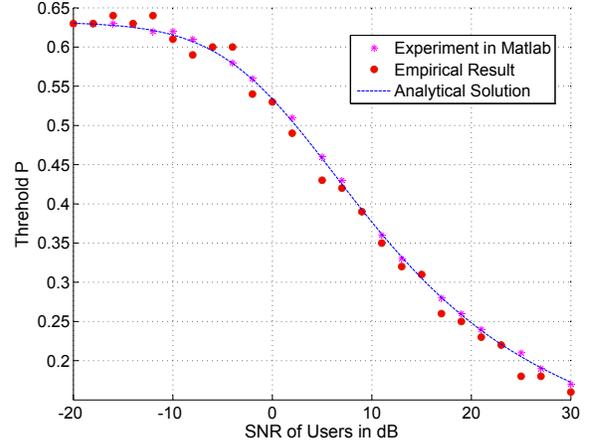}
  \end{center}
  \vspace{-130pt}
 \caption{Comparison of analytical and experimental results}\label{Fig: Optimzation_analytical}
  %\vspace{-100pt}
\end{figure}

Finally we present comparison between optimum p* values calculated by above equation to those obtained by simulation of full system with 100 users in figure ( \ref{Fig: Optimzation_analytical} )

%\begin{figure}[h]
%\centering
%%\includegraphics[scale=0.6]{Through_put_100_50_users.pdf}
%\resizebox {8 cm}{8 cm}{\includegraphics[width=\textwidth]{fig_optimizationforSNR.pdf}}
%\vspace*{-80pt}
%\caption{Comparison of analytical and experimental results}\label{Fig: Optimzation_analytical}
%\end{figure}

\section{Simulation Results}

\begin{figure}[h]
  %\vspace{-20pt}
  \begin{center}
    \includegraphics[width=0.5\textwidth]{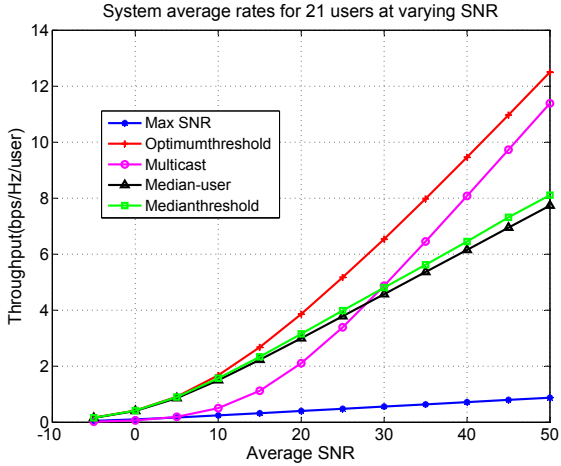}
  \end{center}
  \vspace{-150pt}
  \caption{System throughput for 21 users with equal Avg.SNR }\label{Systemthroughput20users}
  %\vspace{+10pt}
\end{figure}

Figure \ref{Systemthroughput20users} compares the Unicast (equ. (21) in \cite{Multi_vs_Uni}), Multicast  ( equ. (21) in  \cite{Multi_vs_Uni} ), median-user \cite{Praveen_Kumar} , median-threshold and optimized threshold scheduler throughputs for increasing values of Average SNR. The number of users is set as 21 and average SNR changes from -5dB to 50dB in steps of 5dB. From the figure, we can see that system throughputs of all scheduling algorithms increase with increasing average SNR. The system throughput of Unicast increases slowly as it only chooses one user to receive packets all the time. Whereas Multicast gains as the number of user’s increases, but when average SNR is low, it performs worse than Unicast. The median-threshold performs better than median user scheduler. The proposed optimized threshold scheme adjusts threshold of scheduling according to average SNR, hence achieves better throughput.
 
\begin{figure}[h!]
  %\vspace{-10pt}
  \begin{center}
    \includegraphics[width=0.5\textwidth]{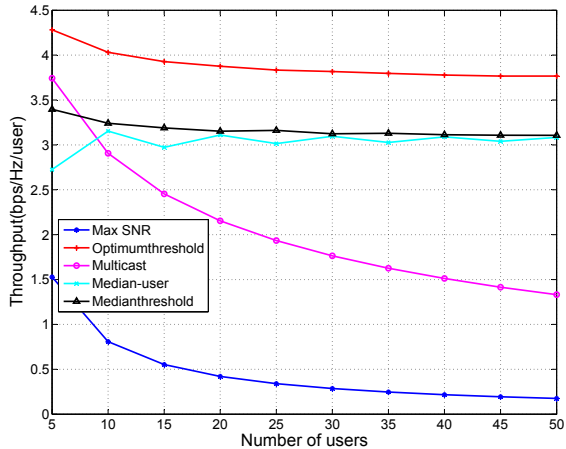}
  \end{center}
  \vspace{-10pt}
  \caption{System throughput for varying number of users all at 20dB}\label{Systemthroughput20dB}
  %\vspace{-100pt}
\end{figure}

In figure \ref{Systemthroughput20dB}, the average SNR of all users is set as 20 dB, and the number of users is change from 5 to 50 in step of 5. We can see that the Unicast and Multicast schemes get poor performance because the former exploits only multiuser diversity gain and the latter only the broadcast gain. Both decrease as the number of user’s increases. The median user scheduling algorithm has much higher system throughput as it offers both the gains partially. The proposed median case of the optimal threshold scheduler shows a slightly better performance compared to median user, the reason being that even though the median converges as the number of users increases the rates is dependent on the number of users always 

%%\begin{figure}[h!]
%  %\vspace{-10pt}
%  \begin{center}
%    \includegraphics[width=0.5\textwidth]{./Fig_Systemthroughput20dB.pdf}
%  \end{center}
%  \vspace{-150pt}
%  \caption{System throughput for varying number of users all at 20dB}\label{Systemthroughput20dB}
%  %\vspace{-100pt}
%\end{figure}

\section{Conclusion}
A new threshold based scheduler is presented. This scheduler tries to exploit both mulituser diversity and multicasting gain simultaneously. Median Threshold Scheduler (MTS) as the name suggests uses median of a distribution as threshold. MTS delivers higher throughput than Unicast and Multicast over wide range of SNR regimes. However Multicast outperforms MTS in High ($>20dB$) SNR regimes.\\

When threshold value p* is 0 threshold based scheduler becomes Multicast and when p* is 1 it converges to Unicast scheduler. We presented a closed form solution to optimum p* given an average SNR value. Threshold scheduler based on p* is termed as Optimum Threshold Scheduler (OTS). OTS outperforms Mulicast and Unicast over all SNR regimes. It converges to Multicast as SNR tends to very large values.
For a practical SNR range of -30 to 50 dB   $0.1<p*<0.6$  i.e. even in high SNR regimes Multicast is suboptimal. \\

OTS always has a minimum rate which is independent of number of users and based only on average SNR values. We would like to highlight that minimum rate of OTS is better than Unicast or Multicast over wide range of SNR values. Such a minimum rate is useful for determining QoS (minimum bitrate) requirements, and as it is independent of number of users which means given, an avg SNR one can find minimum bit rate irrespective of number of users in system.

In future work, we would study performance of proposed scheduler with delay as a performance metric. And also consider throughput-delay trade off as in \cite{Praveenkumar_throuputdelay}

In heterogeneous networks Mutlicast is absolutely fair that is it delivers same rate to all users. Unicast in such case is greedy scheduler and is biased towards users with high SNR values.  A similar threshold based strategy can be used here as an intermediate solution between Multicast and Unicast.

\nocite{OMS}
\appendix

\section{System Lower Bound}
\label{Appendix: lower bound}
We select user for transmission when its instantaneous SNR $\gamma \ge {{\gamma }_{th}}$ .
Hence, we have 
\[{{\gamma }_{th}}\le \gamma \]
Where  ${{\gamma }_{th}}$ is set threshold and $\gamma $   is instantaneous SNR at which scheduler transmits data\\
As log is a monotonically increasing function hence we can write 
\begin{equation} \label{equ:logth le loggamma}
{{\log }_{2}}\left( 1+{{\gamma }_{th}} \right)\le {{\log }_{2}}\left( 1+\gamma  \right)
\end{equation}

Multiplying on both sides by bandwidth $W>0$ we obtain

\begin{equation} \label{equ:rth le rsys}
{{R}_{th}}\le {{R}_{system}}
\end{equation}

Furthermore as the rate Rth is obtained only when users pass the threshold average Rth is given by 
\[\left( 1-p \right)\times {{R}_{th}}\] and as 
\[0<\left( 1-p \right)\le 1\]
We have
\begin{equation} \label{equ:rth 1-p  le rsys}
\left( 1-p \right)\times {{R}_{th}}\le {{R}_{th}} 
\end{equation}

From \ref{equ:rth le rsys} and \ref{equ:rth 1-p  le rsys}
 we have  
\begin{equation}
\left( 1-p \right)\times {{R}_{th}} \le {{R}_{system}}
\end{equation}

${{\gamma}_{th}}$ used in ( \ref{equ:logth le loggamma} ) is obtained by using quantile or inverse CDF function given by 
\begin{equation}  \label{equ:quantile}
{{\gamma }_{th}}=-\ln \left( 1-p \right)\overline{\gamma } 
\end{equation}
where p is quantile/probability  chosen\\ 

From ( \ref{equ:quantile} ) and ( \ref{equ:rth le rsys} ) Lower bound of the system is given by  
\begin{equation} \label{equ:system lim}
{{R}_{low}}=(1-p){{log}_{2}}(1-\ln (1-p))W 
\end{equation}
 
\vspace{25pt}

\section{Derivation of system optimum values}
\label{Appendix:optimzation}
\textbf{\emph{Derivation of system optimum values:}}

\vspace{25pt}

We wish to find optimum value of p which maximizes system lower bound ${{R}_{low}}$, ( \ref{equ:system lim} )\\

Hence, we have following optimization problem 
\[p*=\operatorname*{arg\,max}_p (1-p)\log 2(1-\ln (1-p))W\] 
\[{s.t.}\;\;\;\;{    p>0}\]
\[\;\;\;{  \& }   \;\;\;\;   {p<1}\] 
We can re write objective function substituting $x=1-p$ 
\[f\left( x \right)=x\frac{\ln (1-\ln (x)\overline{\gamma })}{\ln \left( 2 \right)}W\] 
Dropping constants we have objective function as 
\[f\left( x \right)=x\ln \left( 1-\ln \left( x \right)\overline{\gamma } \right)\] 
This is an optimization problem with inequality constrains
Hence we can solve it using KKT theorem.
We find KKT multipliers u* for functions g(x*) which satisfy following conditions

\begin{enumerate}

\item $\mu *\ge 0$
\item $\nabla f\left( x \right)+{{\left( \mu * \right)}^{T}}\nabla g\left( x* \right)=0$
\item $\mu {{*}^{T}}g(x*)={{0}^{T}}$
\end{enumerate}

It can be easily shown that above conditions are satisfied when $\mu *=0$ 
Hence condition 2 can be written as
\[\nabla f\left( x \right)=\ln \left( 1-\ln \left( x \right)\bar{\gamma } \right)+\frac{-\bar{\gamma }x}{\left( 1-\ln \left( x \right)\bar{\gamma } \right)}{{\times }^{1}}{{/}_{x}}=0\]
Substituting 
\begin{equation} \label{equ:sub t}
t=\left( 1-\ln \left( x \right)\overline{\gamma } \right)
\end{equation}
\[\ln t-\overline{\frac{\gamma }{t}}=0\] 
\[\therefore t\ln t=\overline{\gamma } \]
Which we can rewrite as 
\[\left( \ln t \right){{e}^{\left( \ln t \right)}}=\overline{\gamma }\] 
Finally we substitute $\ln t=z$ 
\[z{{e}^{z}}=\overline{\gamma }\] 
Solution to this equation is given by principal root of Lambert Function $W\left( x \right)$ \cite{Lambert_W}
\[z=W\left( \overline{\gamma } \right)\] by definition of Lambert Function 
\[\ln t=W\left( \overline{\gamma } \right)\] 
\[t={{e}^{W\left( \overline{\gamma } \right)}}\] 
From \ref{equ:sub t}
\[\left( 1-\ln \left( x \right)\overline{\gamma } \right)={{e}^{W\left( \overline{\gamma } \right)}}\] 
Rearranging
\[\ln x=\frac{1}{\overline{\gamma }}-\frac{{{e}^{W\left( \overline{\gamma } \right)}}}{\overline{\gamma }}\] 

\[x=e^{\frac{1}{{\overline{\gamma }}}-\frac{e^{W({\overline{\gamma }})}}{{\overline{\gamma }}}}\]
putting $x=1-p$
\begin{equation} \label{equ: optimal p}
p*=\frac{{{e}^{\frac{{{e}^{W(\overline{\gamma })}}}{\overline{\gamma }}}}-\sqrt[\overline{\gamma }]{e}}{{{e}^{\frac{{{e}^{W(\overline{\gamma })}}}{\overline{\gamma }}}}}
\end{equation}

\nocite{Tse04fundamentalsof}

\bibliographystyle{ieeetr}
\bibliography{abc}

\begin{thebibliography}{1}

\bibitem{Knoop_and_Humblet}
R.~Knopp and P.~Humblet, ``Information capacity and power control in
  single-cell multiuser communications,'' in {\em Communications, 1995. ICC '95
  Seattle, 'Gateway to Globalization', 1995 IEEE International Conference on},
  vol.~1, pp.~331 --335 vol.1, jun 1995.

\bibitem{Praveenkumar_throuputdelay}
P.~Gopala and H.~El~Gamal, ``On the throughput-delay tradeoff in cellular
  multicast,'' in {\em Wireless Networks, Communications and Mobile Computing,
  2005 International Conference on}, vol.~2, pp.~1401 -- 1406 vol.2, june 2005.

\bibitem{Praveen_Kumar}
P.~K. Gopala and H.~Gamal, ``Opportunistic multicasting,'' in {\em Signals,
  Systems and Computers, 2004. Conference Record of the Thirty-Eighth Asilomar
  Conference on}, vol.~1, pp.~845 -- 849 Vol.1, nov. 2004.

\bibitem{OMS}
T.~ping Low, M.~on~Pun, Y.-W. Hong, and C.-C. Kuo, ``Optimized opportunistic
  multicast scheduling (oms) over wireless cellular networks,'' {\em Wireless
  Communications, IEEE Transactions on}, vol.~9, pp.~791 --801, february 2010.

\bibitem{Lambert_W}
R.~Corless, G.~Gonnet, D.~Hare, D.~Jeffrey, and D.~Knuth, ``On the lambert;
  function,'' {\em Advances in Computational Mathematics}, vol.~5,
  pp.~329--359, 1996.
\newblock 10.1007/BF02124750.

\bibitem{Multi_vs_Uni}
N.~El~Heni and X.~Lagrange, ``Multicast vs multiple unicast scheduling in
  high-speed cellular networks,'' in {\em Vehicular Technology Conference,
  2008. VTC Spring 2008. IEEE}, pp.~2456 --2460, may 2008.

\bibitem{Tse04fundamentalsof}
D.~Tse and P.~Viswanath, ``Fundamentals of wireless communications,'' 2004.

\end{thebibliography}

\end{document}